# Interlayer Electronic Coupling on Demand in a 2D Magnetic Semiconductor

Nathan P. Wilson[1,*], Kihong Lee[2,*], John Cenker[1,*], Kaichen Xie[3,*], Avalon H. Dismukes[2], Evan J. Telford[2,4], Jordan Fonseca[1], Shivesh Sivakumar[3], Cory Dean[4], Ting Cao[3,†], Xavier Roy[2,†], Xiaodong Xu[1,3,†], Xiaoyang Zhu[2,†]

[1] Department of Physics, University of Washington, Seattle, WA 98195 USA
[2] Department of Chemistry, Columbia University, New York, NY 10027 USA
[3] Department of Material Science & Engineering, University of Washington, Seattle, WA 98195 USA
[4] Department of Physics and Astronomy, Columbia University, New York, NY 10027 USA

[*]These authors contributed equally to this work.
[†]Correspondence should be addressed to: tingcao@uw.edu (TC); xr2114@columbia.edu (XR); xuxd@uw.edu (XX); xyzhu@columbia.edu (XYZ).

**ABSTRACT. When monolayers of two-dimensional (2D) materials are stacked into van der Waals structures, interlayer electronic coupling can introduce entirely new properties, as exemplified by recent discoveries of moiré bands that host highly correlated electronic states and quantum dot-like interlayer exciton lattices. Here we show the magnetic control of interlayer electronic coupling, as manifested in tunable excitonic transitions, in an A-type antiferromagnetic 2D semiconductor CrSBr. Excitonic transitions in bilayer and above can be drastically changed when the magnetic order is switched from layered antiferromagnetic to the field-induced ferromagnetic state, an effect attributed to the spin-allowed interlayer hybridization of electron and hole orbitals in the latter, as revealed by GW-BSE calculations. Our work uncovers a magnetic approach to engineer electronic and excitonic effects in layered magnetic semiconductors.**

One of the most exciting prospects of 2D materials is their stacking into natural or artificial van der Waals (vdW) structures, in which interlayer electronic coupling can result in new and emergent properties. For example, interlayer hybridization turns the momentum direct bandgaps in transition metal dichalcogenide (TMDC) monolayers to indirect bandgaps in bilayers or multilayers[1,2], introduces moiré bands to host highly correlated phenomena in twisted graphene[3,4] and TMDC bilayers[5–7], and results in the formation of periodic arrays of potential traps for excitons in TMDC heterobilayers[8–11]. In these examples, interlayer hybridization is pre-determined and only in limited cases can it be tuned, e.g., with mechanically rotatable structures [12], with hydrostatic pressure [13], and with vertical electric fields [14–17]. Here we explore in-situ tuning of interlayer electronic hybridization based on the control of layered magnetic order. A-type antiferromagnets are ideal materials for this purpose: they consist of vdW ferromagnetic (FM) monolayers that are coupled antiferromagnetically along the stacking direction. Such interlayer antiferromagnetic (AFM) order can be switched to FM with an external magnetic field, often accompanying a change of properties and symmetry. Tuning the spin structures of layered antiferromagnets has thus led to a number of emerging physical phenomena. Examples based on 2D $CrI_3$ include giant tunneling magneto-resistance via spin filtering effects[18], very large second harmonic generation (SHG) from the AFM state due to magnetic state induced inversion symmetry breaking[19], and tuning inelastic light scattering and spin waves via symmetry controls[20]. In the layered magnet $MnBi_2Te_4$, a topological quantum phase transition occurs when the intrinsic interlayer AFM state is fully polarized to the FM state with an external magnetic field[21,22].

In this work, we report the unique magneto-electronic coupling effects that emerge in CrSBr, a 2D material that combines a direct electronic bandgap with layered A-type AFM order[23]. We examine this coupling by probing optical transitions associated with Wannier excitons that are sensitive to interlayer electronic coupling in CrSBr. The use of Wannier excitons as signatures of electronic-magnetic coupling in CrSBr contrasts with $CrI_3$ where the dominant optical transitions are from localized and parity forbidden d-d orbitals centered on the Cr atom[24]. The lattice of CrSBr consists of vdW layers made of two buckled planes of CrS terminated by Br atoms (Fig. 1a). These layers stack along the *c* axis to produce an orthorhombic structure with $P_{mmn}$ ($D_{2h}$) space group. The mechanical exfoliation of CrSBr single crystals produces elongated flakes, a manifestation of the anisotropic structure of the material (Fig. 1b). The long and short lateral edges of the crystals correspond to the crystallographic *a* and *b* axes determined by single crystal X-ray diffraction (see *methods*). Below the bulk Néel temperature ($T_N$ ~132 K), each CrSBr vdW layer orders ferromagnetically and couples antiferromagnetically to adjacent layers. CrSBr possesses biaxial magnetic anisotropy, with easy and intermediate magnetic axes along the crystallographic *b*- and *a*-axes, respectively, and a hard axis along $c^{23}$. A recent SHG study confirmed that this magnetic structure persist to the FM monolayer and AFM bilayer[25]. Unlike other 2D magnets, the Néel temperature in CrSBr is found to increase with decreasing layer number, from $T_N$ = 132 K in the bulk to $T_N$ = 150 K in the bilayer[25].

Electronic structure calculations of monolayer CrSBr in its FM ground state within the GW approximation reveal a semiconducting bandgap of ~1.8 eV and highly anisotropic band dispersion; see the GW band structure of bilayer Fig. 1c and DFT band structures of one, two, three layers in Fig. S1. The valence band maximum (VBM) is at the Γ point, and two nearly degenerate conduction band minima (CBM) appear at Γ and X points. Despite a different quasiparticle bandgap and dispersion, our GW calculations that include the self-energy corrections are in agreement with previous DFT calculations that also predicted anisotropic and spin-polarized bands [26–28]. Away from Γ, there is considerable anisotropy of the conduction bands, with significant dispersion along Γ–Y and almost flat bands along Γ–X. This feature is consistent with the calculated optical matrix element that is dipole-allowed along the *b*-axis, but forbidden along the *a*-axis, for interband transitions from the VBM to the CBM at the Γ point (further discussions of the optical selection rules can be found in supplementary Fig. S1d). Similar to those in TMDCs [29], the dominant optical transitions come from Wannier excitons in CrSBr (see below), with the calculated exciton binding energy of ~ 0.5 eV for a free-standing monolayer.

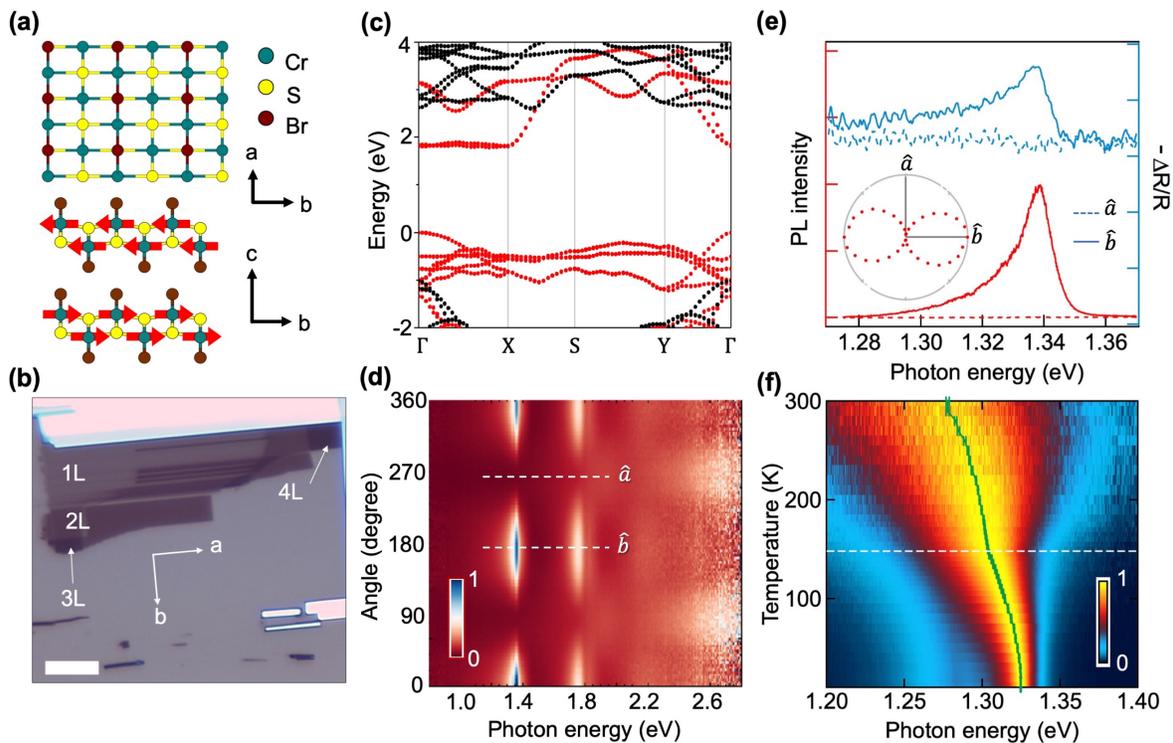

**Figure 1 | Structure and optical properties of CrSBr**. **a**, Crystal and magnetic structures of CrSBr. Top image shows a top-view of a single layer and the bottom image shows a side-view of a bilayer in which the AFM order is represented by the red arrows. **b**, Optical microscope image of exfoliated CrSBr. Scale bar is 5 μm. **c**, Calculated quasiparticle band structure of monolayer CrSBr. The bands of majority- and minority-spin electrons are shown in red and black, respectively. **d**, Polarization resolved differential reflectance spectra of bilayer CrSBr. White dashed lines represent polarization along $\hat{a}$ and $\hat{b}$ axes. **e**, Comparison of differential reflectance spectra (blue, right axis) and PL spectra (red, left axis) with polarization along the $\hat{b}$ (solid) and $\hat{a}$ (dashed) axis, respectively, from bilayer CrSBr. The inset shows PL intensity vs. polarization angle. **f**, PL spectra from bilayer CrSBr as a function of temperature. The spectrum at each temperature is intensity normalized to its maximum. The green dots are peak positions. The white dashed line shows $T_N$ (for bilayer CrSBr) which coincides with a change in PL peak shape. Measurements in **d** and **e** are carried out at a sample temperature of T = 5 K.

The anisotropic optical transition is captured in polarization-resolved optical spectroscopy on exfoliated CrSBr flakes. Fig. 1d shows the linear polarization-resolved differential reflectance (ΔR/R) spectra of bilayer (2L) CrSBr. The lowest-energy exciton peak at ~1.34 eV, exhibits near-perfect linear dichroism with a strong resonant response along the *b*-axis and no detectable resonance along *a*. This behavior is mirrored in the photoluminescence (PL) spectra. The bright PL emission at 1.34 eV is observed with near-unity linear polarization (red curves in Fig. 1e) and this optical anisotropy is observed for all thicknesses of CrSBr flakes, regardless of magnetic order (Fig. S2a-g). A comparison in Fig. 1e of the polarization-resolved PL spectra (red) and differential reflectance spectra (blue) underlines the excellent agreement between the position of the excitonic peaks, which is consistent with a direct bandgap transition.

Evidence for the electronic-magnetic order coupling is first found in the temperature-dependent PL spectra. For monolayer (1L) CrSBr (Fig. S3a,e), the PL spectrum only shows a gradual blueshift of the emission with decreasing temperature as expected for a direct-gap semiconductor. For bilayer CrSBr (Fig. 1f), the blueshift is accompanied by a change in the peak shape across the magnetic phase transition, as evidenced by a kink at 150 ± 5 K, which closely match the bilayer $T_N$ determined by SHG[25]. The magneto-electronic coupling becomes even more evident for thicker flakes (Fig. S3b-d, f-g) as we discuss below.

To gain a deeper understanding of the magneto- electronic coupling in CrSBr, we measured PL as a function of applied magnetic field (*B*) at 5 K and establish that the magneto-excitonic coupling originates from interlayer electronic interaction. For monolayer CrSBr, the PL spectra show no discernible evolution as *B* is swept along the easy (Fig. 2a) or hard axes (Fig. S4a). In stark contrast to the monolayer results, the PL response of the bilayer varies dramatically with *B*. Fig. 2b presents bilayer PL spectra as *B* is swept along the easy axis. The PL changes abruptly at $B_c = 0.134 \pm 0.003$ T and is otherwise constant above and below this transition (see also Fig. S4b). A comparison of monolayer and bilayer PL spectra at selected B fields along the *b* axis can be found in Fig. S5. Field sweeps along the intermediate and hard axes result in a continuous evolution of the PL responses up to saturation fields $B_{sat}$ of ~0.9 and ~1.6 T, respectively, beyond which the PL spectra remain unchanged (Fig. 2c-d, see also Fig. S4c and S4d). This spin-canting process is corroborated by the reflectance magnetic circular dichroism (RMCD) measurements with *B* field along the c-axis (Fig. 2e). The $B_{sat}$ values for bilayer measured by RMCD and inferred from magneto-PL are identical (Fig. 2c). Similar switching (along *b*) and canting (along *a* or *c*) behaviors with external magnetic field are also observed in the differential reflectance spectra of the bilayer (Fig. S6).

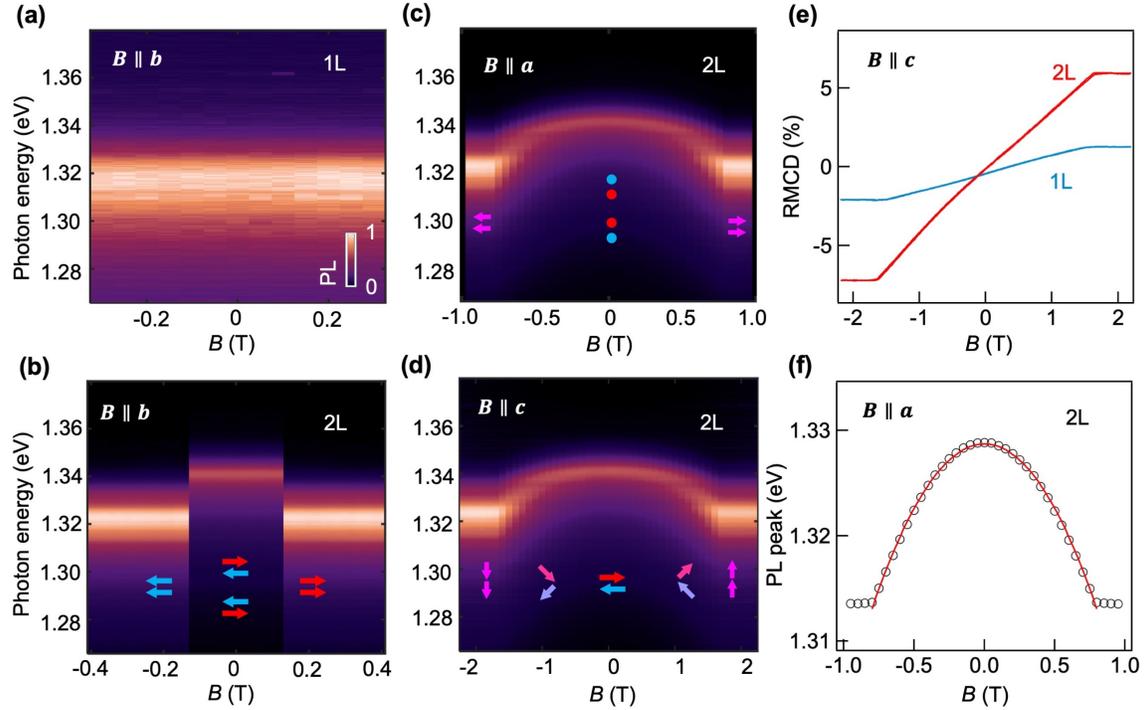

**Fig. 2 | Excitons coupled to magnetic order. a,** Magnetic field dependence of monolayer PL spectrum with the field oriented along the easy axis. **b, c, d,** Magnetic field dependence of bilayer PL spectrum along the easy, intermediate, and hard axes, respectively. **e,** RMCD field sweeps of monolayer (blue) and bilayer (red) CrSBr with field along the c axis. The excitation energy is 1.959 eV. The linear RMCD field dependence with abrupt saturation suggests spin canting behavior, characteristic of anisotropic magnetism. **f,** Field dependent PL peak position (center-of-mass, circles) for bilayer CrSBr as a function of magnetic field along the *a* axis. Red curve: quadratic fit to experimental data below saturation field.

The above results demonstrate that the dramatic change of exciton properties arises from the tuning of interlayer magnetic order. When *B* is along the easy axis, the abrupt switch in the PL spectra of the bilayer at $B_c$ comes from a spin flip transition (i.e. transition from AFM to FM order), resulting in a sudden transformation of the electronic structure and excitonic transitions. When $B$ is along the intermediate or hard axes, spin canting produces continuous changes to the interlayer magnetic order, concurrent with the continuous evolution of the electronic structure and the PL spectra. As we show below, a GW-BSE calculation reproduces the redshift of the exciton from AFM to FM, and a perturbation picture predicts a quadratic energy shift in $B$ below $B_{sat}$, as confirmed in Fig. 1f for spin canting when $B$ is applied to the intermediate axis.

We use first-principles GW-BSE calculations (details in Methods) to obtain the quasiparticle band structures of bilayer CrSBr in the AFM (Fig. 3a) and FM (Fig. 3b) states. In the AFM bilayer, the product symmetry of time reversal and spatial inversion makes the band structure degenerate in spins. In each Bloch band near the doubly degenerate CBM and VBM, the spin-up or spin-down electrons are localized at the top and bottom layer, respectively, since their interlayer hybridization is suppressed by the interlayer AFM order. In the FM bilayer, by contrast, the electrons in the two layers can resonantly couple with each other, leading to band splitting of the CBM and VBM and a band gap reduction of ~ 0.1 eV relative to the AFM bilayer.

The GW-BSE calculations unveil the nature of the optical transitions. In the AFM bilayer, the bright excitons in the top and bottom layers are virtually decoupled due to the anti-aligned spins between layers. The lowest energy bright excitons are two-fold degenerate (with energy difference < 0.5 meV), and account for the optical transitions at ~1.34 eV in the experimental spectra. A comparison between the experimentally measured transition energy and the calculated exciton excitation energy (~1.23 eV) shows a ~ 0.1 eV difference, which arises from the error of GW-BSE calculations (details in method). The calculated exciton wavefunction (top view in Fig. 3c) extends over several unit cells, and is about 2 ~ 3 times more delocalized along the $b$ axis than along the $a$ axis, revealing an anisotropic Wannier character. The sideview of the exciton wavefunction in figure 3d confirms that the electron is localized in the same layer as the hole. In the FM bilayer, by contrast, the bright excitons in the two layers are no longer decoupled states due to the interlayer hybridization of the electron and hole orbitals. For the lowest-energy exciton in the FM bilayer, when the hole is fixed in one layer, the associated electron shows significant amplitude in both layers (shown in Fig. 3e with a typical side view). We further quantify interlayer spatial distribution of the exciton wavefunction by first integrating the wavefunction module square in the top or bottom layer, with the hole fixed to a given position in the bottom layer (Fig. 3d and 3e), and then summed over all the possible hole positions in the bottom layer (see SI for details). The sum of the exciton wavefunction module square in the top layer is < 1% of that in the bottom layer in the AFM bilayer, and ~ 50% of that in the bottom layer in FM bilayer. In other words, the probability density to find the bound electron in the other layer than the fixed hole is > 50 times larger in the FM state than the AFM state. The calculated optical spectra in the AFM and FM states are shown in figure 3f. By turning on the interlayer electronic coupling in the FM state, the most visible effect is a redshift of the optically bright exciton by ~10 meV from that in the AFM state. This calculated redshift is consistent with the experimental measurements, as shown by the differential reflectance spectra from bilayer CrSBr in the AFM (blue) and field-induced FM (red) states, respectively (Fig. 3h). The same redshift is seen in the magnetic field dependent PL spectra in figures. 2b and 2e.

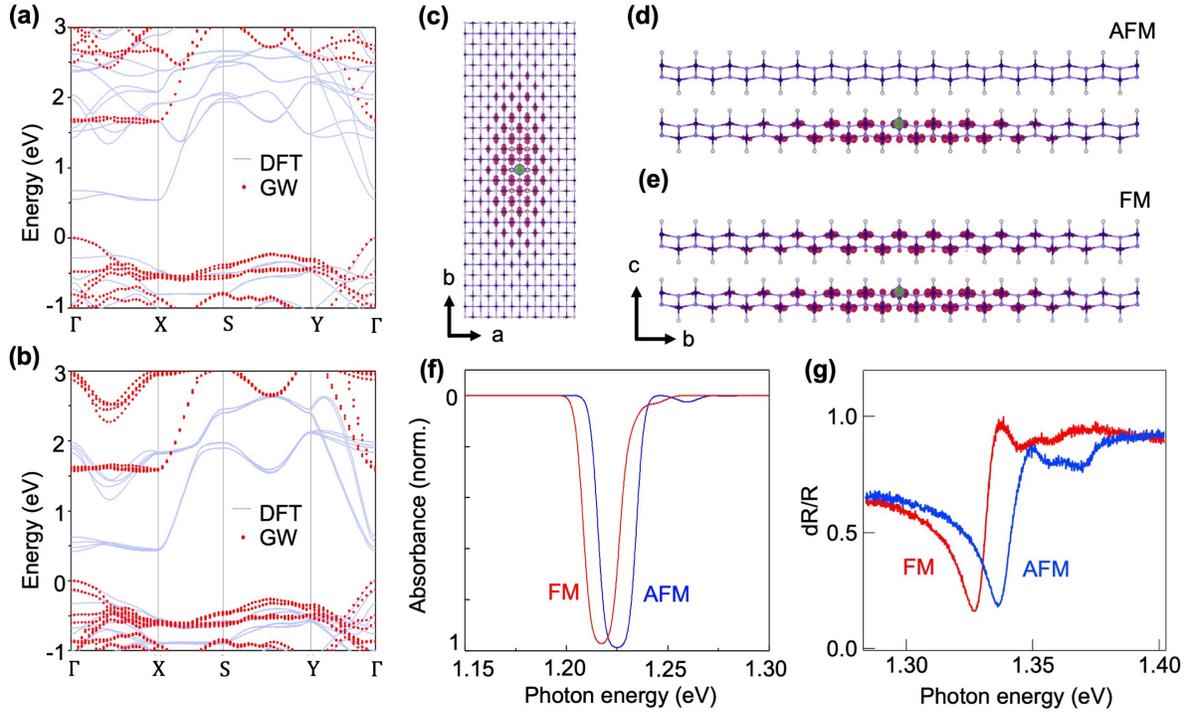

**Figure 3. Magnetic order-dependent band structure and excitonic transitions.** (a, b) Band structures of the AFM (a) and FM (b) CrSBr bilayers. Blue and red dots are Kohn-Sham band energies calculated using the DFT-PBE method and quasiparticle band energies calculated using the GW method, respectively. In the AFM bilayer, bands are degenerate in spin. In the FM case, the bands of majority spins are shown (those of minority spins have much larger gaps). (c, d, e) Real-space wavefunction of the lowest-energy exciton. Plots show the wavefunction module square of a bound electron with the hole fixed near a Cr atom (labelled by a green circle) in the bottom layer. The iso-surfaces represent the amplitude module square set at 1% of its maximum. (c) top view in the AFM bilayer. Top view in the FM bilayer is very similar. (d) side view in the AFM bilayer showing that the electron is virtually localized in the same layer as the hole, and (e) side view in the FM bilayer showing that the electron wavefunction is delocalized across both layers for a hole (green circle) fixed in the bottom layer. (f) Optical absorption spectra of linearly polarized light (polarization along *b* axis) of the AFM (blue) and FM (red) CrSBr bilayers. A Gaussian broadening of 0.005 eV is used to model the imaginary part of the dielectric function. For polarization along *a* axis, absorbance goes to 0. (g) Experimentally measured differential reflectance spectra of the AFM (blue) and FM (red) CrSBr bilayers.

The interlayer hybridization explains the stark difference between the discrete switching behavior of the excitonic transitions when the magnetic field is along the easy axis (Fig. 2b) and the continuous evolution (Fig. 2c,d) when it is along the intermediate/hard axis. The wavefunctions near the VBM or CBM of each CrSBr layer can be approximated by the product of the spatial and the spinor parts, with the spatial part nearly independent of the spin orientation. The interlayer hopping integral $t_h$ is therefore proportional to the inner product of the spinor wavefunctions of adjacent layers, $t_h \propto \langle S_1|S_2\rangle = \cos(\theta/2)$, where $\theta$ is the angle between the magnetization vectors of the layers. For the spin flip transition, $\theta$ jumps from $\pi$ to 0 at $B_c$, while for the spin canting behavior, $M = M_{sat}\cos(\theta/2) \propto B$ up to $B_{sat}$. When $B$ is along the intermediate axis, lowest-order energy shift in the perturbation theory gives, $\Delta E_B = B \cdot M \propto B^2$, as is confirmed in Fig. 2f. Moreover, the interlayer hybridization across the vdW interface makes relaxation into the lowest energy bright exciton more efficient and, thus, more competitive with non-radiative recombination. This explains the abrupt increase in PL intensity across the spin flip transition (Fig 2b) and the gradual increase in PL intensity as the spins are progressively canted away from the easy axis (Fig. 2c,d).

Our interpretation implies that thicker layers should give rise to intermediate excitonic transitions due to transitional magnetic states between the AFM and the fully polarized FM orders. To identify these different magnetic states, we performed PL and differential reflectance measurements as a function of $B$ on 3-layer (3L) and 4-layer (4L) CrSBr flakes. We focus on 4L in Fig. 4 and the 3L data is summarized in figure S7. The temperature dependence of the PL spectrum of 4L CrSBr (Fig. 4a) shows that the broad PL peak observed at high temperature resolves into two narrower branches below $T_N$ = 139 K. This splitting is consistent with our understanding: the excitons are strongly coupled between different layers in the high temperature paramagnetic phase due to interlayer hybridization, and become localized in individual layers in the AFM phase. We expect the layer-localized excitons to experience different environments, such as local dielectric screening depending on the layer they reside, and the asymmetric environment may also brighten otherwise forbidden optical transitions, e.g., involving the second conduction band ~40 meV higher in energy (see Fig. S1a-d). These may provide an explanation for the PL peak splitting observed in the AFM phase for three layer or thicker CrSBr flakes.

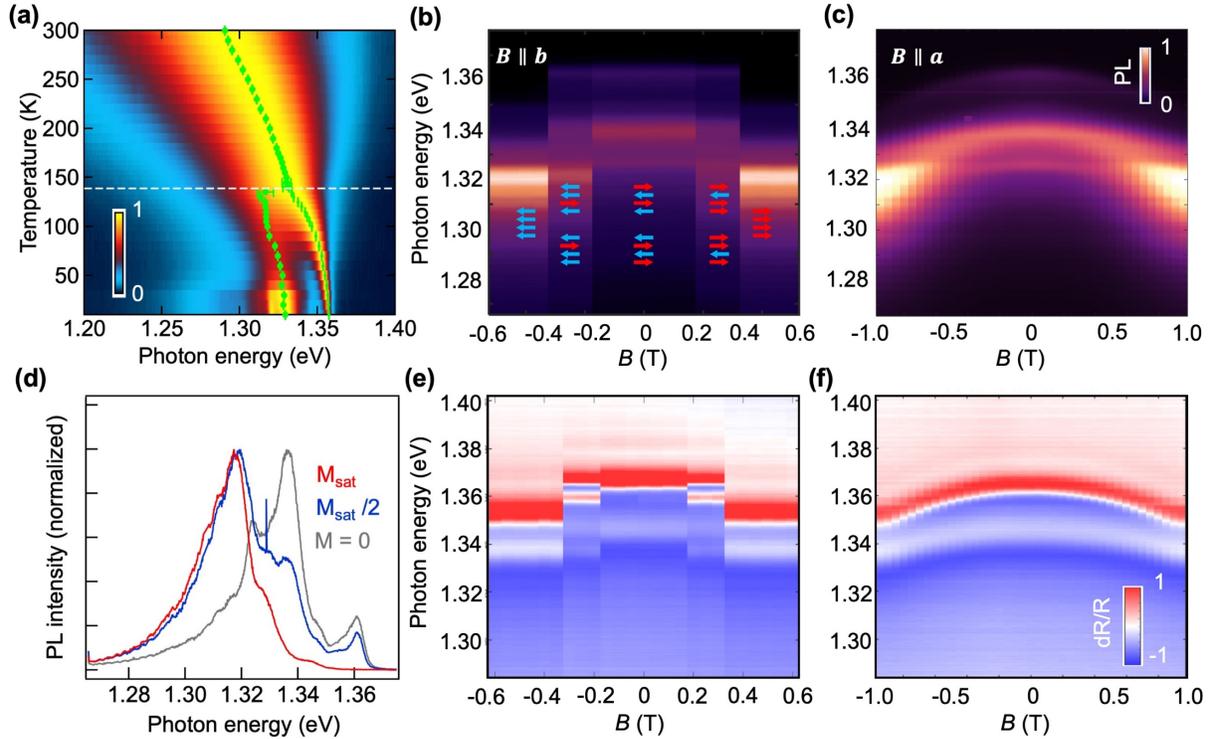

**Fig. 4 | Excitons in 4-layer CrSBr. a**, Peak normalized PL spectrum of 4L CrSBr as a function of temperature, with the Néel temperature marked by a white dashed line. Below the transition temperature, a second PL branch emerges. The green symbols are peak positions. **b**, Magneto-PL of 4L CrSBr along the easy axis, with possible magnetic configurations denoted by the red/blue arrows. Layer-flip transitions occur at 0.17 T and 0.33 T. **c**, **b**, Magneto-PL of 4L CrSBr along the intermediate axis. Instrument limitation for in-plane magnetic field of 0.95 T prevents us from reaching $B_{sat}$ along the intermediate axis. **d,** Comparison of PL spectra in the three different magnetic configuration of 4L CrSBr. The half-magnetized state resembles a superposition of the AFM and FM configuration. **e, f,** Magneto differential reflectance spectra of 4L CrSBr along the easy axis (e) and intermediate axis (f), respectively.

Measuring the PL and differential reflectance spectra as a function of $B$ at 5 K provides additional insights into the coupling between magnetic order and excitonic transitions. When $B$ is swept along the easy axis, the PL spectra of 4L CrSBr change abruptly at certain critical fields, with no detectable changes between those fields (Fig. 4b). Unlike the single magnetic transition observed for the bilayer, there are two separate transitions at $B_c$ = 0.17 and 0.33 T for 4L CrSBr. These two transitions correspond to spin flip transitions of individual layers, as illustrated by the arrows in figure 4b and similarly observed before in a few layer CrI$_3$ [30]. The PL response when $B$ is along the intermediate axis (Fig. 4c) mimics that of the bilayer (Fig. S8 presents PL data when $B$ is along the hard axis). The PL peaks redshift with increasing B and the higher energy peak vanishes when approaching $B_{sat}$.

As in bilayer CrSBr, the PL intensity of the 4L flake increases abruptly when the AFM state undergoes layer-by-layer spin flip transitions, or gradually when the spins are gradually canted towards the fully polarized FM state. Figure 4d compares intensity-normalized PL spectra at zero ($M = 0$), half ($M_{sat}/2$), and fully saturated magnetization ($M_{sat}$) along the easy axis. The PL spectrum at $M_{sat}/2$ resembles a superposition of those at $M = 0$ and $M_{sat}$, and can be understood as containing the emission responses from both AFM and FM interfaces. Similar to the PL results, the $B$ dependence of the differential reflectance spectra of 4L CrSBr displays clear optical signatures of intermediate magnetic states, spin flipping transitions when $B$ is along the easy axis (Fig. 4e) and spin canting behavior when $B$ is along the intermediate axis (Fig. 4f). This agreement between PL and differential reflectance spectra in all magnetic states is a further proof of strong magneto-excitonic coupling in the direct-gap semiconductor.

The results presented here demonstrate an effective approach to tailor interlayer electronic coupling in vdW semiconductors by control of their magnetic orders. The interlayer electronic coupling, as revealed in excitonic transitions, can be used as an "on/off" or "rotary" switch when an external magnetic field is applied to the easy axis of an AFM bilayer or multilayer, and as a "dimmer" when the field is applied to the intermediate or hard axis. Moreover, the magneto-electronic coupling may enable simple optical means to probe or manipulate spin information, such as the launching or tracking of spin waves. Finally, the possibility of controlling interlayer twist angles in artificially stacked magnetic semiconductor bilayers or multilayers adds rich dimensions to the burgeoning field of moiré physics, with the tantalizing prospect of controlling periodic arrays of inter-layer hybridization and moiré bands by spin order and external magnetic fields.

**Methods:**

<u>Sample preparation and characterization</u>. We synthesized single crystal CrSBr using a modified procedure based on ref. [31], as detailed elsewhere[25,32]. The long axis of bulk needle crystals of CrSBr has been correlated to the *a* crystal axis by X-ray diffraction (XRD) experiments. Bulk crystals were exfoliated by first cleaving them on tape with a known, fixed crystal orientation, then transferring them mechanically on an Si/SiO$_2$ substrate passivated by 1-dodecanol[25]. The orientation of the crystals on the tape transfers to the exfoliated crystals, allowing for the crystal orientation of exfoliated samples to be correlated to prior XRD characterization. Samples were handled, transported, and characterized entirely in an oxygen and moisture-free environment (both less than 1 ppm).

Optical spectroscopy. Magnetic field dependence measurements were carried out in a custom Montana Instruments closed-cycle cryostat equipped with a 3-axis vector magnet. Sample crystal axes were aligned visually to the vector magnet axes to within 2° along all directions. RMCD measurements were performed using non-resonant 633 nm laser light with quarter-wave modulation from a photoelastic modulator, focused to a spot size of ~ 2 μm on the sample by an aspheric singlet lens. Unless otherwise indicated, PL and reflectance measurements were carried out at a sample temperature of 5±1 K. Excitation for magneto-PL measurements was from a 633 nm laser at a power of 10 μW using a beam spot size of ~ 1 μm. For magneto-reflectance spectroscopy, a thermal white light source was employed. Spectra were measured using a CCD array following dispersal by a 500 mm spectrometer. Temperature-dependent PL experiment was performed with a 633 nm HeNe laser as excitation source and on-sample power of 200 μW. Emission was collected with a Princeton Instruments PyLoN-IR and SpectraPro HRS-300.

First-Principles Calculations. The mean-field starting point of the GW calculations uses density functional calculations (DFT) within the spin-polarized generalized gradient approximation (GGA), performed using the Quantum ESPRESSO package[33]. We employed norm-conserving pseudopotentials, with a plane-wave energy cutoff of 85 Ry. In the structural relaxation, we included dispersion corrections within the D2 formalism to account for the van der Waals interactions[34]. The structure was fully relaxed until the force on each atom was smaller than 0.01 eV/Å. The calculated lattice constants along the *a* and *b* axes are 3.5 Å and 4.7 Å, respectively, in agreement with experimental results[32]. The calculated interlayer distance is 8.1 Å in bilayer. The scalar-relativistic and full-relativistic band structures show little differences near the VBM or CBM. The GW[35] calculations were carried out using the BerkeleyGW package[36] at the $G_0W_0$ level. The supercell in the monolayer and bilayer calculations uses out-of-plane lattice constants of 16 Å and 28 Å. A truncated Coulomb interaction is employed along the out-of-plane direction to avoid interactions between the free-standing CrSBr layers and its periodic images. In the calculation of the electron self-energy, the dielectric matrix was constructed with a cutoff energy of 35 Ry. The dielectric matrix and the self-energy were calculated on an $8 \times 6 \times 1$ k-grid. 10 subsampling points along the in-plane diagonal of the supercell are included in the calculation of the dielectric function[37]. A static remainder approach is used, together with 1,700 bands in the bilayer calculation[38]. These parameters lead to a converged quasiparticle bandgap within 0.1 eV. The exciton energy levels and wavefunctions are calculated using the GW-BSE methods[39]. The exciton interaction kernel is calculated on a 32 x 24 x 1 k-grid in bilayer, which converges the exciton binding energy to within 0.1 eV, similar to a previous report on $CrI_3$.[40]

**Acknowledgements:** The temperature-dependent PL measurements were supported by the US Air Force Office of Scientific Research (AFOSR) grant FA9550-18-1-0020 (to X.-Y.Z., and X. R.). Magneto-optical spectroscopy measurements are mainly supported by the DoE, BES under award DE-SC0018171. Synthesis, structural characterization, and polarization resolved photoluminescence measurements of CrSBr is supported by the Center on Programmable Quantum Materials, an Energy Frontier Research Center funded by the U.S. Department of Energy (DOE), Office of Science, Basic Energy Sciences (BES), under award DE-SC0019443. Computational resources were provided by Hyak at UW. JF and KX acknowledges the Graduate Fellowship from Clean Energy Institute funded by the State of Washington. T.C. acknowledges support from the Micron Foundation. XYZ acknowledges partial support for laser equipment by the Vannevar Bush Faculty Fellowship through Office of Naval Research Grant # N00014-18-1-2080.

**Author contributions:**
XYZ, XX, KL, and NPW conceived this work. Bulk crystals were synthesized and characterized by AHD and EJT with supervision by XR and CD. Sample preparation was carried out by KL, and AHD, assisted by NPW and JC. Temperature dependent measurements were performed by KL with supervision from XYZ. Field-dependent and polarization-dependent measurements were performed by NPW and JC with supervision from XX. The vector magnet was operated by JF. KX, SS, and TC performed first-principles calculations that interpreted the results. The manuscript was prepared by NPW, KL, JC, KX, TC, XX, and XYZ in consultation with all other authors. All authors read and commented on the manuscript.

**References:**

1. Mak, K., Lee, C., Hone, J., Shan, J. & Heinz, T. Atomically Thin MoS2: A New Direct-Gap Semiconductor. *Phys. Rev. Lett.* **105**, 136805 (2010).

2. Splendiani, A. *et al.* Emerging photoluminescence in monolayer MoS2. *Nano Lett.* **10**, 1271–1275 (2010).

3. Cao, Y. *et al.* Unconventional superconductivity in magic-angle graphene superlattices. *Nature* **556**, 43–50 (2018).

4. Cao, Y. *et al.* Correlated insulator behaviour at half-filling in magic-angle graphene superlattices. *Nature* **556**, 80–84 (2018).

5. Wang, L. *et al.* Correlated electronic phases in twisted bilayer transition metal dichalcogenides. *Nat. Mater.* **19**, 861–866 (2020).

6. Tang, Y. *et al.* Simulation of Hubbard model physics in $WSe_2/WS_2$ moiré superlattices. *Nature* **579**, 353–358 (2020).

7. Regan, E. C. *et al.* Mott and generalized Wigner crystal states in $WSe_2/WS_2$ moiré superlattices. *Nature* **579**, 359–363 (2020).

8. Seyler, K. L. *et al.* Signatures of moiré-trapped valley excitons in $MoSe_2/WSe_2$ heterobilayers. *Nature* **567**, 66–70 (2019).

9. Tran, K. *et al.* Evidence for moiré excitons in van der Waals heterostructures. *Nature* **567**, 71–75 (2019).

10. Jin, C. *et al.* Observation of moiré excitons in WSe2/WS2 heterostructure superlattices. *Nature* **567**, 76–80 (2019).

11. Alexeev, E. M. *et al.* Resonantly hybridized excitons in moiré superlattices in van der Waals heterostructures. *Nature* **567**, 81–86 (2019).

12. Ribeiro-Palau, R. *et al.* Twistable electronics with dynamically rotatable heterostructures. *Science* **361**, 690–693 (2018).

13. Yankowitz, M. *et al.* Tuning superconductivity in twisted bilayer graphene. *Science* **363**, 1059–1064 (2019).

14. Chen, G. *et al.* Signatures of tunable superconductivity in a trilayer graphene moiré superlattice. *Nature* **572**, 215–219 (2019).

15. Jauregui, L. A. *et al.* Electrical control of interlayer exciton dynamics in atomically thin heterostructures. *Science* **366**, 870–875 (2019).

16. Shimazaki, Y. *et al.* Strongly correlated electrons and hybrid excitons in a moiré heterostructure. *Nature* **580**, 472–477 (2020).

17. Tang, Y. *et al.* Tuning layer-hybridized moiré excitons by the quantum-confined Stark effect. *Nat. Nanotechnol.* **16**, 52–57 (2021).

18. Song, T. *et al.* Giant tunneling magnetoresistance in spin-filter van der Waals heterostructures. *Science* **360**, 1214–1218 (2018).

19. Sun, Z. *et al.* Giant nonreciprocal second-harmonic generation from antiferromagnetic bilayer CrI3. *Nature* **572**, 497–501 (2019).

20. Huang, B. *et al.* Tuning inelastic light scattering via symmetry control in the two-dimensional magnet CrI3. *Nat. Nanotechnol.* **15**, 212–217 (2020).

21. Liu, C. *et al.* Quantum phase transition from axion insulator to Chern insulator in MnBi2Te4. *arXiv Prepr. arXiv1905.00715* (2019).

22. Li, J. *et al.* Magnetically controllable topological quantum phase transitions in the antiferromagnetic topological insulator MnBi 2 Te 4. *Phys. Rev. B* **100**, 121103 (2019).

23. Göser, O., Paul, W. & Kahle, H. G. Magnetic properties of CrSBr. *J. Magn. Magn. Mater.* **92**, 129–136 (1990).

24. Seyler, K. L. *et al.* Ligand-field helical luminescence in a 2D ferromagnetic insulator. *Nat. Phys.* **14**, 277–281 (2018).

25. Lee, K. *et al.* Magnetic Order and Symmetry in the 2D Semiconductor CrSBr. *arXiv Prepr. arXiv2007.10715* (2020).

26. Guo, Y., Zhang, Y., Yuan, S., Wang, B. & Wang, J. Chromium sulfide halide monolayers: intrinsic ferromagnetic semiconductors with large spin polarization and high carrier mobility. *Nanoscale* **10**, 18036–18042 (2018).

27. Wang, C. *et al.* A family of high-temperature ferromagnetic monolayers with locked spin-dichroism-mobility anisotropy: MnNX and CrCX (X= Cl, Br, I; C= S, Se, Te). *Sci. Bull.* **64**, 293–300 (2019).

28. Wang, H., Qi, J. & Qian, X. Electrically tunable high Curie temperature two-dimensional ferromagnetism in van der Waals layered crystals. *Appl. Phys. Lett.* **117**, 83102 (2020).

29. Wang, G. *et al.* Colloquium: Excitons in atomically thin transition metal dichalcogenides. *Rev. Mod. Phys.* **90**, 21001 (2018).

30. Huang, B. *et al.* Layer-dependent ferromagnetism in a van der Waals crystal down to the monolayer limit. *Nature* **546**, 270 (2017).

31. Beck, J. Über Chalkogenidhalogenide des Chroms Synthese, Kristallstruktur und Magnetismus von Chromsulfidbromid, CrSBr. *Zeitschrift für Anorg. und Allg. Chemie* **585**, 157–167 (1990).

32. Telford, E. J. *et al.* Layered Antiferromagnetism Induces Large Negative Magnetoresistance in the van der Waals Semiconductor CrSBr. *Adv. Mater.* **32**, 2003240 (2020).

33. Giannozzi, P. *et al.* QUANTUM ESPRESSO: a modular and open-source software project for quantum simulations of materials. *J. Phys. Condens. Matter* **21**, 395502 (2009).

34. Grimme, S. Semiempirical GGA-type density functional constructed with a long-range dispersion correction. *J. Comput. Chem.* **27**, 1787–1799 (2006).

35. Hybertsen, M. S. & Louie, S. G. Electron correlation in semiconductors and insulators: Band gaps and quasiparticle energies. *Phys. Rev. B* **34**, 5390 (1986).

36. Deslippe, J. *et al.* BerkeleyGW: A massively parallel computer package for the calculation of the quasiparticle and optical properties of materials and nanostructures. *Comput. Phys. Commun.* **183**, 1269–1289 (2012).

37. Felipe, H., Qiu, D. Y. & Louie, S. G. Nonuniform sampling schemes of the Brillouin zone for many-electron perturbation-theory calculations in reduced dimensionality. *Phys. Rev. B* **95**, 35109 (2017).

38. Deslippe, J., Samsonidze, G., Jain, M., Cohen, M. L. & Louie, S. G. Coulomb-hole

summations and energies for G W calculations with limited number of empty orbitals: A modified static remainder approach. *Phys. Rev. B* **87**, 165124 (2013).

39. Rohlfing, M. & Louie, S. G. Electron-hole excitations and optical spectra from first principles. *Phys. Rev. B* **62**, 4927 (2000).

40. Wu, M., Li, Z., Cao, T. & Louie, S. G. Physical origin of giant excitonic and magneto-optical responses in two-dimensional ferromagnetic insulators. *Nat. Commun.* **10**, 1–8 (2019).

Interlayer Electronic Coupling on Demand in a 2D Magnetic Semiconductor

Nathan P. Wilson[1,*], Kihong Lee[2,*], John Cenker[1,*], Kaichen Xie[3,*], Avalon H. Dismukes[2], Evan J. Telford[2,4], Jordan Fonseca[1], Shivesh Sivakumar[3], Cory Dean[4], Ting Cao[3,†], Xavier Roy[2,†], Xiaodong Xu[1,3,†], Xiaoyang Zhu[2,†]

[1] Department of Physics, University of Washington, Seattle, WA 98195 USA
[2] Department of Chemistry, Columbia University, New York, NY 10027 USA
[3] Department of Material Science & Engineering, University of Washington, Seattle, WA 98195 USA
[4] Department of Physics and Astronomy, Columbia University, New York, NY 10027 USA

[*]These authors contributed equally to this work.

[†]Correspondence should be addressed to: tingcao@uw.edu; xr2114@columbia.edu; xuxd@uw.edu; xyzhu@columbia.edu.

**Contents:**

Supplementary figure 1: Electronic properties of layered CrSBr

Supplementary figure 2: Layer-dependent excitons with optical anisotropy

Supplementary figure 3: Temperature dependent PL

Supplementary figure 4: Magneto-PL of monolayer and bilayer CrSBr

Supplementary figure 5: Magneto-PL of monolayer and bilayer CrSBr at selected B fields

Supplementary figure 6: Bilayer magneto-reflectance spectroscopy

Supplementary figure 7: Trilayer magneto-PL

Supplementary figure 8: Four-layer hard axis magneto-PL

Supplementary method: Integral of exciton wavefunction module square

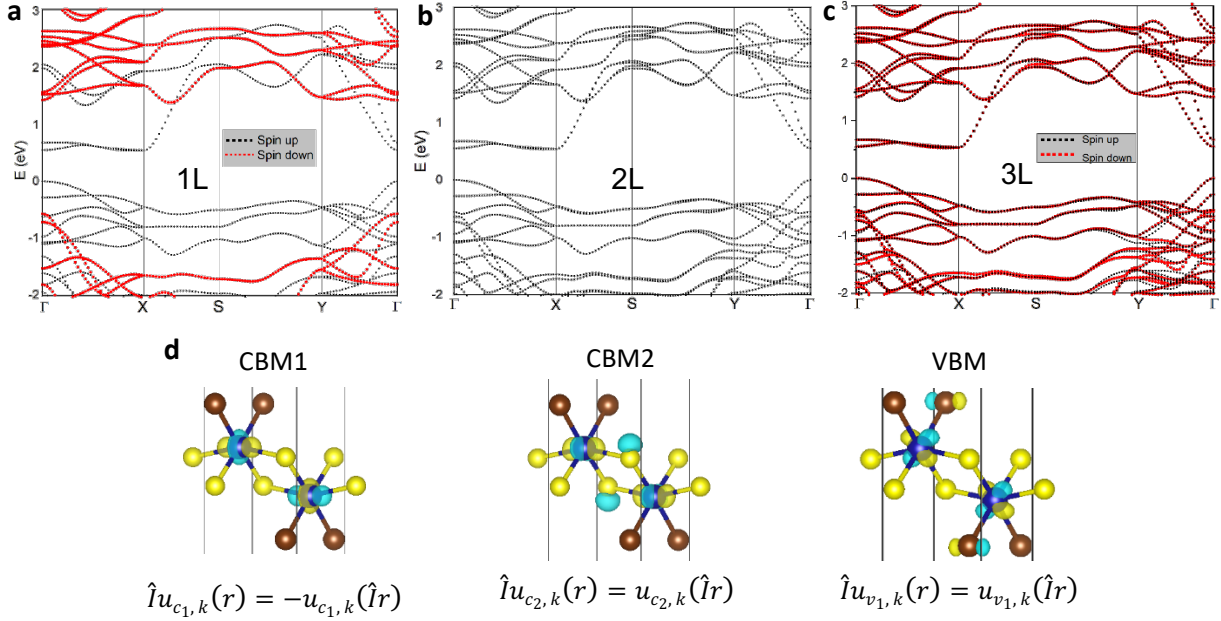

**Figure S1 | Electronic properties of layered CrSBr**
**a-c,** DFT band structures *without GW correction* for 1-3 layer CrSBr, respectively, in their AFM states. Majority spins are shown in black, and minority spins in red in odd-numbered layers. In the case of the even-numbered layers such as the bilayer, spin-up and spin-down bands are degenerate. Due to this electronic decoupling, there is no significant thickness dependence to the calculated band structure. We note the presence of a second conduction band with a local minimum at Γ (CBM2), which is approximately 40 meV higher in energy than the global conduction band minimum (CBM1) at Γ in the GW calculations depicted in main text fig. 1c. **d,** Diagrams depicting the orbital composition of CBM1, CBM2, and the valence band maximum (VBM). The transformation of the wavefunction $u_{i,k}(r)$ under spatial inversion ($\hat{I}$) is noted below each diagram. Based on the parity and spin of the band extrema, the transition between VBM and CBM1 is allowed, but the transition between VBM and CBM2 is dipole forbidden. However, by reduction of symmetry (e.g. by an asymmetric dielectric environment), the VBM to CBM2 transition could be brightened, offering a possible explanation for the presence of multiple exciton resonances near 1.35 eV in the CrSBr photoluminescence and reflectance spectra (as shown in main text figures 3 and 4, and in the supplementary figures that follow).

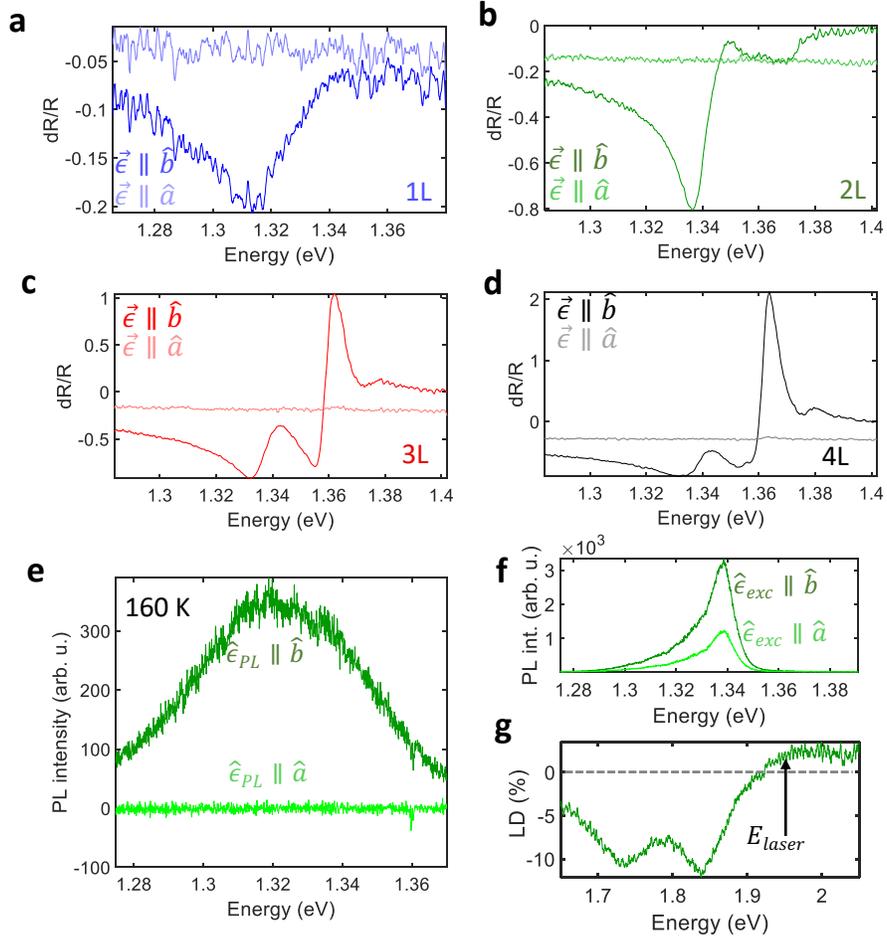

**Figure S2 | Layer-dependent excitons with optical anisotropy**
**a-d,** Polarization-resolved differential reflectance spectra of 1-4 L CrSBr in the vicinity of the lowest energy exciton resonance. All observed resonances are fully polarized along the b direction. **e,** Polarization-resolved PL of bilayer CrSBr above the magnetic transition temperature. The degree of polarization remains near 100%, showing that the polarization is not connected to magnetic order. We note the emergence of multiple strong resonances in multilayer samples and refer to the main text for discussion of these new spectral features. **f,** PL excitation polarization dependence, with PL collected along the b-axis and the excitation polarization along the a- and b-axis, demonstrating the existence of an excitation selection rule. The degree of excitation polarization dependence is around 50%, far greater than the degree of linear dichroism (LD) (**g**) at the laser energy.

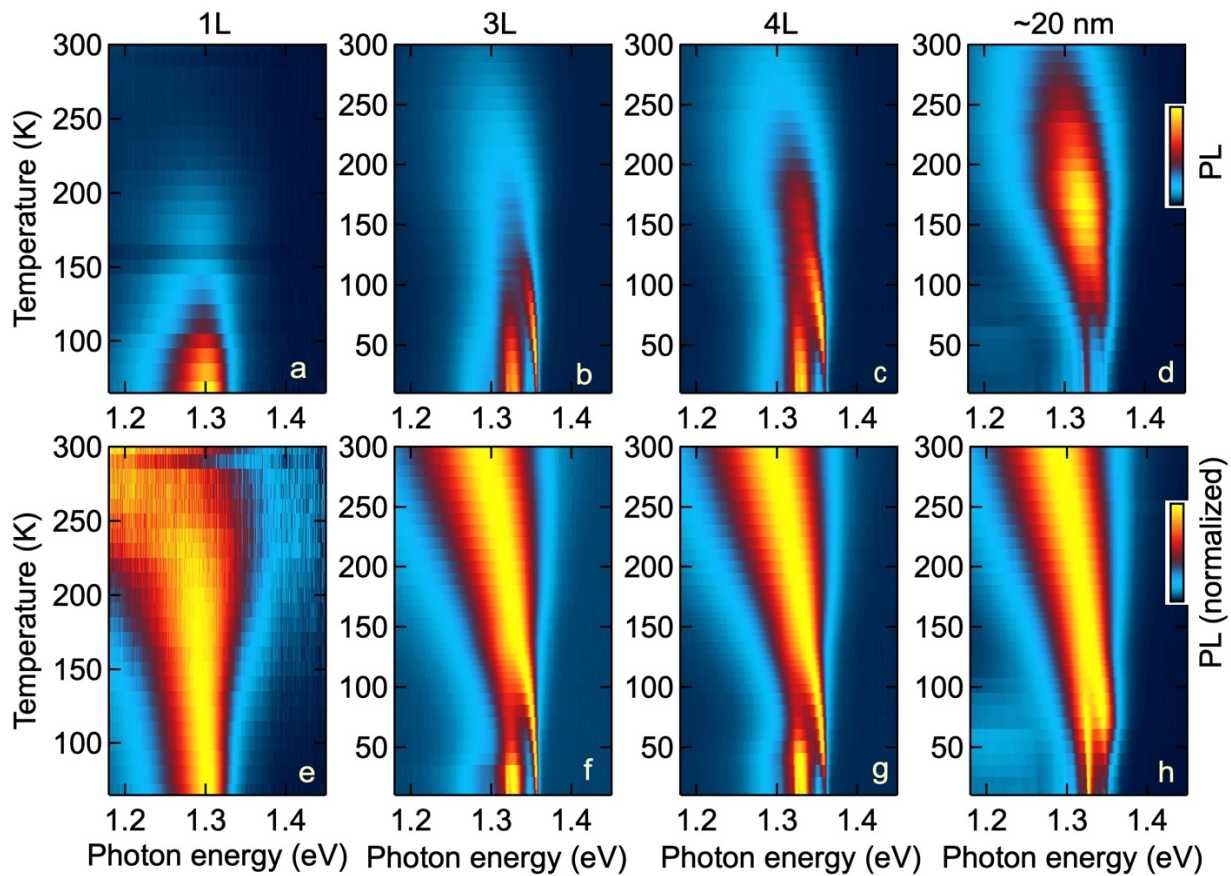

**Figure S3 | Temperature dependent PL**
Temperature-dependent photoluminescence (PL) spectra from a) 1L, b) 3L, c) 4 L, d) thin bulk CrSBr flakes. The bottom panels show corresponding PL spectra with peak intensity normalized at each temperature: e) 1L, f) 3L, g) 4 L, h) thin bulk CrSBr flakes.

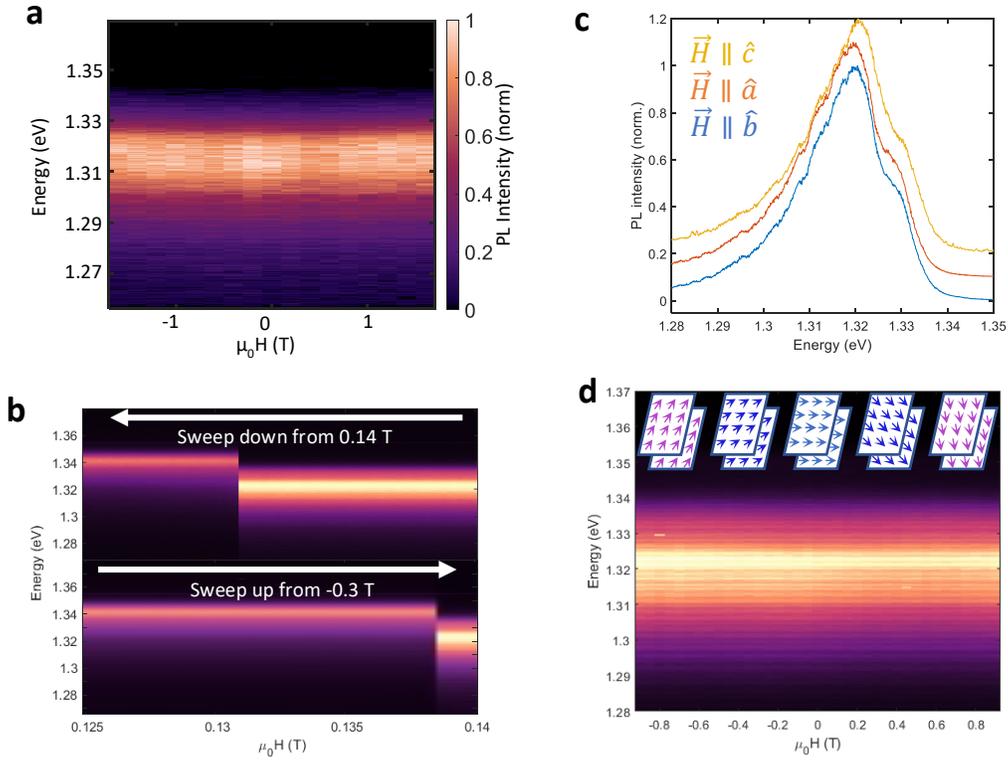

**Figure S4 | Magneto-PL of monolayer and bilayer CrSBr**
**a,** Monolayer PL as a function of magnetic field along the hard axis, showing no response. **b,** Bidirectional magneto-PL sweep of bilayer CrSBr. The spin-flip transition shows hysteretic behavior, with slightly different critical fields for the different sweep directions. **c**. A comparison of PL spectra from bilayer CrSBr in the magnetic field induced FM state (blue), with the those form fully canted states at saturation magnetic fields along the intermediate (orange) and hard (yellow) axes. **d,** Bilayer PL as a function of magnetic field along the intermediate axis in the presence of a 0.3 T easy-axis field which initializes the bilayer into a field-induced FM state.

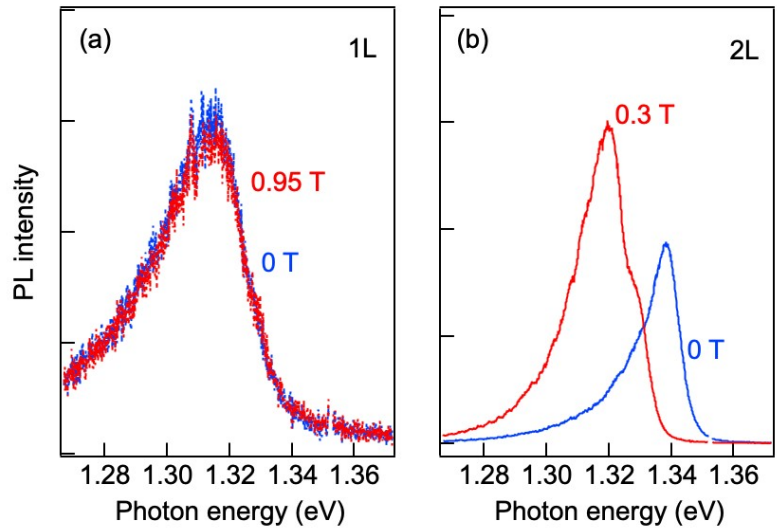

**Figure S5 | Magneto-PL spectra of monolayer and bilayer CrSBr at selected B fields**
Field dependent PL spectra for (a) monolayer and (b) bilayer CrSBr at the indicated magnetic fields along the easy axis. Monolayer: 0.0 T (blue) and 0.95 T (red); Bilayer: 0.0 T (blue) and 0.0.3 T (red).

Note that the PL intensities from the monolayer in (a) are one-order of magnitude lower than those from the bilayer in (b). The PL peak widths from the monolayer are also broader than those from the bilayer. We find that the monolayer sample is more susceptible to degradation with time than the bilayer or thicker samples are, especially exposed to the ambient. We attribute the lower PL intensity and broader peak width in the monolayer as compared to the bilayer to higher defect density in the former.

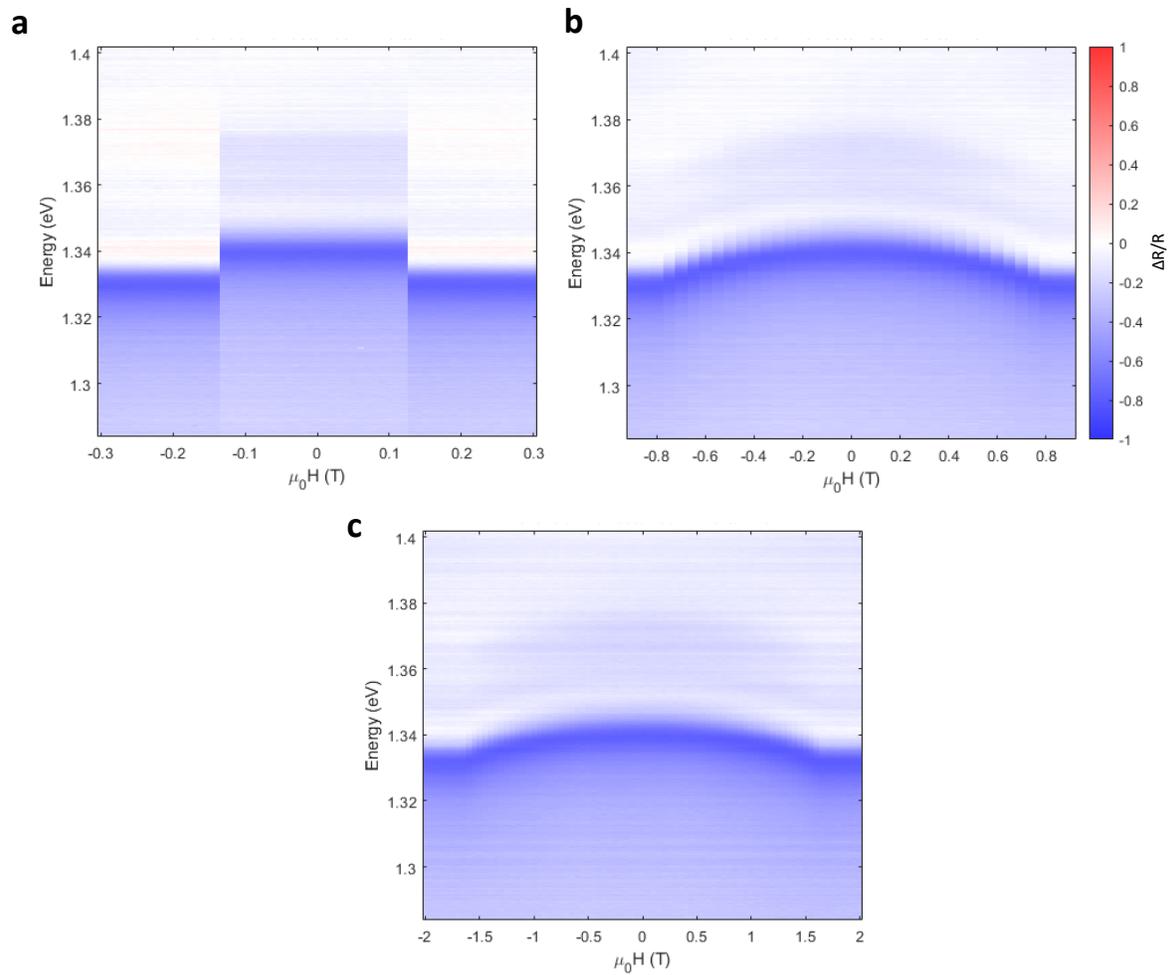

**Figure S6 | Bilayer magneto-reflectance spectroscopy**
**a-c,** Diffeential reflectance of bilayer CrSBr as a function of easy, intermediate, and hard axis magnetic field (respectively).

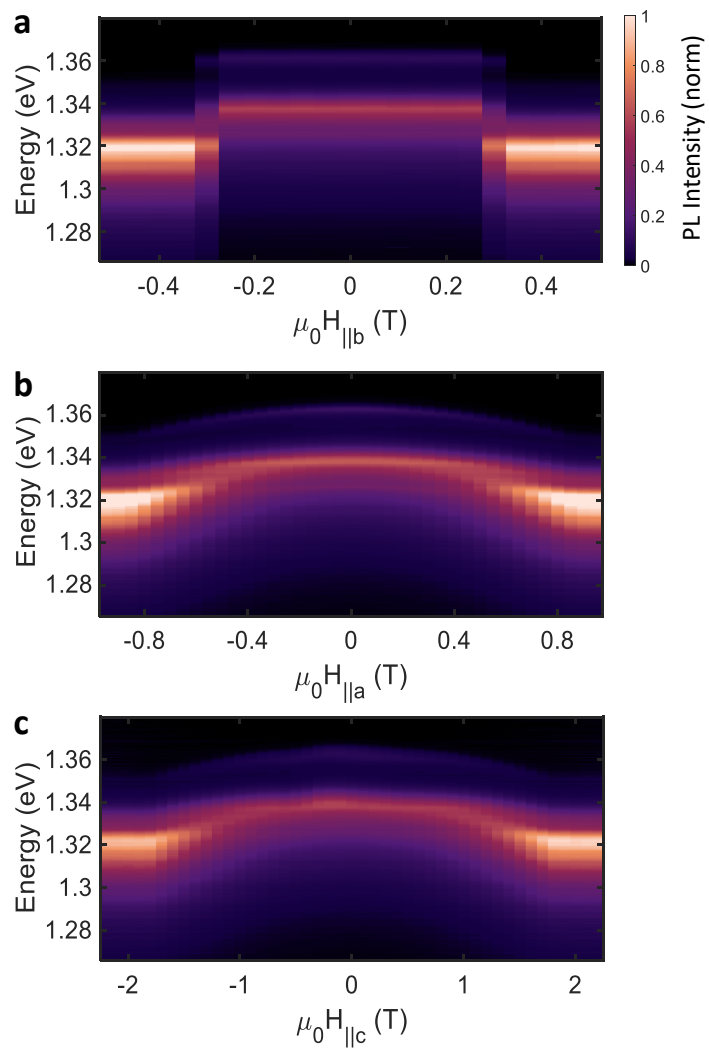

**Figure S7 | Trilayer magneto-PL**
**a-c,** Magneto-PL measurements of trilayer CrSBr with magnetic field along the easy, intermediate, and hard axes, respectively.

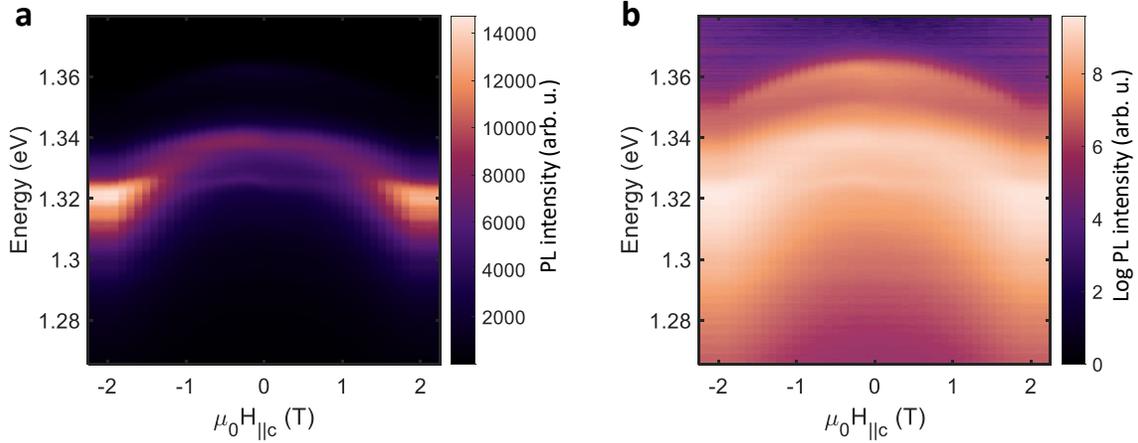

**Figure S8 | Fourlayer hard axis magneto-PL**
**a-b,** Magneto-PL measurement of fourlayer CrSBr with magnetic field along the hard axis. Intensity is plotted in linear and log scales, respectively. The logarithmic scale emphasizes the weaker, higher energy PL features.

Method:

The exciton wavefunction of a selected excited state $|S\rangle$ can be casted into:

$$\Psi^S(\mathbf{r}_e, \mathbf{r}_h) = \sum_{cv\mathbf{k}} A^S_{cv\mathbf{k}} \psi_{c\mathbf{k}}(\mathbf{r}_e) \psi^*_{v\mathbf{k}}(\mathbf{r}_h)$$

In Fig. 3 of the main text, the hole coordinate $\mathbf{r}_h$ is fixed at certain point of the bottom layer, with the electron coordinate $\mathbf{r}_e$ running over a real-space mesh in the supercell. $A^S_{cv\mathbf{k}}$ describes the k-space exciton envelope function for the exciton state $|S\rangle$ in the quasiparticle state representation. $c, v,$ and $\mathbf{k}$ are the conduction-band, valence-band, and k-point indices, respectively. To characterize the spatial distributions of exciton wavefunction with hole fixed in all possible spots in the bottom layer, the integral of the wavefunction module square $\rho^S(\mathbf{r}_e)$ is evaluated by:

$$\rho^S(\mathbf{r}_e) = \int_{\mathbf{r}_h \in B} \Psi^S(\mathbf{r}_e, \mathbf{r}_h) \Psi^{S*}(\mathbf{r}_e, \mathbf{r}_h) d\mathbf{r}_h$$

where $\mathbf{r}_h$ runs over the bottom layer as the location of the hole may result in different electron wavefunction distributions in the FM bilayer. As $\int_{\mathbf{r}_h \in B} \psi^*_{v\mathbf{k}}(\mathbf{r}_h) \psi_{v'\mathbf{k}'}(\mathbf{r}_h) d\mathbf{r}_h = 0, \forall \mathbf{k} \neq \mathbf{k}'$ for two-dimensional materials, $\rho^S(\mathbf{r}_e)$ may be rewritten as:

$$\rho^S(\mathbf{r}_e) = \sum_{cvc'v'\mathbf{k}} A^S_{cv\mathbf{k}} A^{S*}_{cv\mathbf{k}} \psi_{c\mathbf{k}}(\mathbf{r}_e) \psi^*_{c'\mathbf{k}}(\mathbf{r}_e) \int_{\mathbf{r}_h \in B} \psi^*_{v\mathbf{k}}(\mathbf{r}_h) \psi_{v'\mathbf{k}}(\mathbf{r}_h) d\mathbf{r}_h$$

which can be obtained from GW-BSE calculations. Therefore, the ratio $\eta^S$ of the sum of $\rho^S(\mathbf{r}_e)$ in the top layer to that in the bottom layer is used to characterize the layer-resolved spatial distributions of exciton wavefunction :

$$\eta^S = \frac{\int_{\mathbf{r}_e \in T} \rho^S(\mathbf{r}_e) d\mathbf{r}_e}{\int_{\mathbf{r}_e \in B} \rho^S(\mathbf{r}_e) d\mathbf{r}_e}$$

where $\mathbf{r}_e$ runs over the top and bottom layers, respectively. Here the bottom layer refers to the bottom half of the supercell and the top layer refers to the top half of the supercell. The calculated $\eta^S$ is ~ 0.5% in AFM bilayer and ~ 50% in FM bilayer, which confirms that the electron is localized in the same layer as the hole in AFM bilayer and can delocalize over both layers in FM bilayer.